# Impostor Phenomenon Among Software Engineers: Investigating Gender Differences and Well-Being

Paloma Guenes
*Department of Informatics*
*PUC-Rio*
Rio de Janeiro, Brazil
pguenes@inf.puc-rio.br

Rafael Tomaz
*Department of Informatics*
*PUC-Rio*
Rio de Janeiro, Brazil
rtomaz@inf.puc-rio.br

Bianca Trinkenreich
*Department of Computer Science*
*Colorado State University*
Fort Collins, USA
bianca.trinkenreich@colostate.edu

Maria Teresa Baldassarre
*Department of Informatics*
*University of Bari*
Bari, Italy
mariateresa.baldassarre@uniba.it

Margaret-Anne Storey
*Department of Computer Science*
*University of Victoria*
Victoria, Canada
mstorey@uvic.ca

Marcos Kalinowski
*Department of Informatics*
*PUC-Rio*
Rio de Janeiro, Brazil
kalinowski@inf.puc-rio.br

*Abstract*—Research shows that more than half of software professionals experience the Impostor Phenomenon (IP), with a notably higher prevalence among women compared to men. IP can lead to mental health consequences, such as depression and burnout, which can significantly impact personal well-being and software professionals' productivity. This study investigates how IP manifests among software professionals across intersections of gender with race/ethnicity, marital status, number of children, age, and professional experience. Additionally, it examines the well-being of software professionals experiencing IP, providing insights into the interplay between these factors. We analyzed data collected through a theory-driven survey (n = 624) that used validated psychometric instruments to measure IP and well-being in software engineering professionals. We explored the prevalence of IP in the intersections of interest. Additionally, we applied bootstrapping to characterize well-being within our field and statistically tested whether professionals of different genders suffering from IP have lower well-being. The results show that IP occurs more frequently in women and that the prevalence is particularly high among black women as well as among single and childless women. Furthermore, regardless of gender, software engineering professionals suffering from IP have significantly lower well-being. Our findings indicate that effective IP mitigation strategies are needed to improve the well-being of software professionals. Mitigating IP would have particularly positive effects on the well-being of women, who are more frequently affected by IP.

*Index Terms*—Impostor Phenomenon, Imposter Syndrome, Human Aspects, Gender Differences, Software Professionals, WHO-5 Well-Being Index

## I. Introduction

The Impostor Phenomenon (IP), first described by Dr. Pauline Clance in her clinical observations with women [6], refers to a psychological pattern where individuals harbor persistent doubts about their accomplishments and have an internalized fear of being exposed as fraud. IP hinders individuals from realizing their full potential, enjoying their achievements, and pursuing opportunities, as self-doubt and fear undermine their confidence and joy in success [7]. Despite evident success and external validation, those experiencing the Impostor Phenomenon feel their achievements are unearned and attribute them to luck or deception rather than their competence and effort [7]. This phenomenon can lead to chronic self-doubt, anxiety, depression, and burnout [3], [6].

In our previous work, Guenes *et al.* [11] provided the first comprehensive examination of the prevalence of the Impostor Phenomenon among software engineering professionals. Although their research has shown that IP is more prevalent among women [11], there is a lack of studies examining IP across intersectional demographic groups. The intersection of gender and race/ethnicity, for example, in the education domain, showed that Black women's experiences of IP impacted many facets of their professional lives from moments of paralysis to potentially unhealthy over-productivity [9].

Furthermore, while studies in other domains have demonstrated that IP can negatively impact well-being [8], [21], it is important to better understand this relationship in the context of software engineering and provide empirical evidence. A recent study [18] argues that researchers should explore whether gender differences in IP have meaningful implications for individuals' success and well-being. Having a more nuanced understanding of the prevalence of IP and well-being across different demographics can help organizations develop more effective interventions to mitigate IP and improve well-being [19].

This study examines how software professionals experience IP across various intersectional demographics, including combinations of gender with age, experience in software engineering, race/ethnicity, marital status, and number of children. Additionally, it explores the well-being of individuals affected by IP, shedding light on the multifaceted nature of its impact.

## II. Background and Related Work

In this section, we present the theoretical foundation for our research and review the pertinent literature. Initially, we delineate the Impostor Phenomenon with the primary insights derived from our previous study [11], which served as the basis for the current extension. Then, we describe the psychometric instruments utilized for its measurement, along with an overview of the WHO-5 Well-Being Index. Finally, we discuss related work from other areas on the relationship between IP, gender, and well-being.

### A. Impostor Phenomenon in Software Engineers

In 1985, Dr. Clance, one of the psychologists who initially identified the Impostor Phenomenon (IP) in high-achieving women, developed the Clance Impostor Phenomenon Scale (CIPS) to assess the presence and severity of IP [4]. CIPS is a reliable psychometrically validated instrument that has been widely used by researchers and practitioners [14]. It consists of a 20-item questionnaire where each item is rated on a Likert scale from 1 to 5, with 1 being "Not at all true" and 5 being "Very true" [14]. The CIPS score is obtained by summing the responses, a higher score indicates a greater intensity of IP. This score is divided into four scoring categories: 40 or less, representing few impostor characteristics; 41 to 60, representing moderate IP experiences; 61 to 80, representing frequently having impostor feelings; and 80 or more, representing intense IP experiences. In this study, scoring more than 60 means meeting the diagnostic criterion of suffering from IP commonly used in research on IP in Computing [11], [26].

In this paper, we extend our previous research [11], which represents the first comprehensive investigation of the prevalence and manifestation of IP in software engineering professionals in general. In our previous work, we conducted a theory-driven survey that included demographic questions, psychometrically validated instruments to measure IP (CIPS) and well-being (WHO-5), and questions on perceived productivity based on the SPACE [10] productivity measurement framework.

From a diverse group of 624 software professionals, 52.72% were experiencing frequent and intense levels of IP, affecting 48.81% of men and 62.63% of women. These results suggest that IP prevalence differs across gender groups. Furthermore, we also observed disparities across race/ethnicity, and professional roles. Underrepresented groups, including women and black individuals, more frequently suffer from IP than their counterparts. The study also revealed that the prevalence of IP had a statistically significant negative effect on perceived productivity. The scope of the study we reported in [11] did not include exploring the level of IP between different genders concerning other demographic factors, or the analysis of collected data on well-being.

### B. Assessing the Well-Being

The WHO-5 Well-Being Index [17] is a psychometrically validated scale developed by the World Health Organization to measure subjective psychological well-being that is reported to have excellent reliability and construct validity [15]. Comprising five simple, non-invasive questions, it assesses how individuals have felt over the previous two weeks, focusing on their overall mood, energy levels, and interest in daily life. Each item, such as "I have felt cheerful and in good spirits" and "I have felt calm and relaxed", is rated on a scale from 0 (none of the time) to 5 (all of the time), yielding a raw score between 0 and 25. This raw score can be converted to a percentage score by multiplying by four, giving a range from 0 (worst possible quality of life) to 100 (best possible quality of life) [17].

WHO-5 has shown great validity as both a screening tool for depression and as an outcome measure in clinical studies, and has been successfully applied in diverse research fields [15]. Using a WHO-5 cut-off $score \leq 50$ (or a raw score under 13) is recommendable when screening for clinical depression [22], [16]. A WHO-5 $score \leq 28$ is a particularly stringent threshold associated with the level of well-being seen in patients with DSM [1] major depression [22].

### C. Impostor Phenomenon, Gender and Well-being

A study with Canadian students [19] emphasized that individuals with expressive traits, typically associated with stereotypically feminine roles, often report lower autonomy and environmental mastery, which correlates with higher impostor feelings. In contrast, those with instrumental or blended traits tend to experience greater well-being and confidence in their abilities. These findings provide insights into the complex interplay between IP, gender, and well-being, suggesting that the impact of IP on well-being may vary significantly based on one's gender.

More recently, a meta-analytic review further highlights that studying IP and gender is crucial due to significant differences in how men and women experience IP [18]. Women generally score higher on IP measures, which can have implications for mental health, self-esteem, and career advancement [18]. The study also calls for future research to explore nuanced questions, including how IP differs across genders and other inter-related factors. Furthermore, they suggest that researchers should investigate whether gender differences and IP have significant implications for individuals' well-being. This topic will be thoroughly examined in this study.

## III. RESEARCH METHOD

### A. Goal and Research Questions

The research goal of this paper can be stated according to the Goal-Question-Metric paradigm goal definition template [1] as follows; *Analyze* the prevalence of IP in software professionals *with the purpose of* characterizing *with respect to* manifestation across intersections of gender with race/ethnicity, marital status, number of children, age, and software engineering experience; and its relation with well-being *from the point of view* of the researcher *in the context of* software engineering professionals. From this goal, we derived and detailed two Research Questions (RQs):

---

[1] The Diagnostic and Statistical Manual of Mental Disorders (DSM) is the American Psychiatric Association's diagnostic manual for mental disorders.

- **RQ1: How does IP manifest in software professionals across different genders?** To address this question, we performed intersectional analyses of IP by examining its occurrence across combinations of gender with race/ethnicity, marital status, number of children, age, and years of experience in software engineering.
- **RQ2: How does IP affect the well-being of software professionals across different genders?** To address this question, we analyze how IP affects well-being and if there are any observable differences when looking at different genders.

### B. Data Collection

We used the same data collected in the survey previuosly conducted us [11], which is available in our online open science repository[2]. Our data was provided by 624 software professionals and includes demographic information on gender, age, race/ethnicity, country of citizenship, marital status, number of children, and experience. Furthermore, the data comprised the answers to the questions of validated scales to measure IP (20 questions of the CIPS scale) and well-being (5 questions of the WHO-5 scale).

### C. Data Analysis Procedures

To assess the manifestation of IP in software professionals, we calculated frequencies for the prevalence of IP across different genders, considering race/ethnicity, marital status, number of children, age, and experience. The CIPS scale provides a score representing the intensity of IP for each respondent. We categorized the respondents into two groups: those with IP, who scored more than 60 on the CIPS, and those without IP, who scored 60 or less.

For well-being, the WHO-5 scores were multiplied by 4 to standardize the scale, and two distinct ranges were analyzed: 0–50 and 51–100, where scores in the lower range indicate poor mental well-being and may warrant further investigation for potential symptoms of depression. Additionally, the ranges 0–28 and 29–100 were examined, with scores in the lowest range corresponding to well-being levels typically observed in patients with major depression. We applied bootstrapping with confidence intervals to characterize the well-being of software engineering professionals.

To examine the relationship between IP and well-being, we applied the nonparametric Mann-Whitney U-test to check for statistically significant differences (alpha value 0.05) in the WHO-5 scores. This test is appropriate when one variable is nominal (e.g., meeting the IP criterion or not) and the other is ordinal (*e.g.*, the WHO-5 score).

Furthermore, to characterize well-being in software engineering professionals, we calculate confidence intervals using a technique called bootstrapping that has been reported to be more reliable and precise than statistical inferences drawn directly from samples [12], [25]. Bootstrapping calculates confidence intervals by re-sampling our data set, creating many simulated samples to promote a more robust and accurate analysis. Considering our sample size N, to perform bootstrapping, we create new samples S times of the same size N. Re-samples may include a given response zero or more times. We set S to 1000, a value known to yield meaningful statistical results [12]. It is important to note that we applied bootstrapping only for men and women due to the small sample size for other genders.

### IV. RESULTS

In this section, we present the results of our investigation, first describing the population sample and then providing the answers to each research question.

#### A. Study Population

In our previous work [11], we assessed the representativeness of our sample, identifying distribution patterns that confirmed its suitability for reflecting the population of software engineering professionals. The sample consists of 624 responses, including participants of different genders from 26 countries across five continents (see Figure 1). The *Others* category represents participants from other 13 countries.

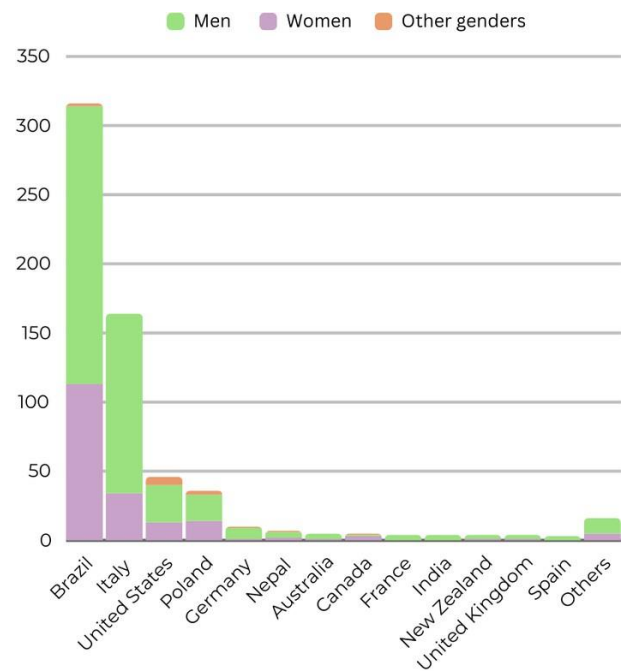

**Fig. 1:** Participants' gender distribution by countries.

As shown in Table I, the sample was primarily composed of men (67.63%), followed by women (30.13%), and other gender identities (2.24%). *Other genders* include participants who identified their gender as non-binary, preferred not to answer, or selected more than one gender. The most common ethnicity in the sample is *White* (78.21%), followed by *Black or African American* (6.89%) and *Asian* (4,49%). *Other* represents *American Indian or Alaska Native*, *Other Race*, and *Prefer not to answer*. The sample also shows a balance between married (49.68%) and single (44.23%) respondents.

---
[2]https://doi.org/10.5281/zenodo.8415205

**TABLE I:** Overview of the sample demographics.

| Survey Question | Options | N | Percentage |
| --- | --- | --- | --- |
| What is your gender? | Man | 422 | 67.63% |
| | Woman | 188 | 30.13% |
| | Non-binary | 7 | 1.12% |
| | Prefer not to answer | 5 | 0.8% |
| | More than one gender | 2 | 0.32% |
| What is your predominant race/ethnicity? | White | 488 | 78.21% |
| | Prefer not to answer | 60 | 9.61% |
| | Black or African American | 43 | 6.89% |
| | Asian | 28 | 4.49% |
| | Other race | 3 | 0.48% |
| | American Indian or Alaska Native | 2 | 0.32% |
| What is your marital status? | Married or Cohabiting | 310 | 49.68% |
| | Single | 276 | 44.23% |
| | Prefer not to answer | 24 | 3.85% |
| | Divorced | 13 | 2.08% |
| | Widow/Widower | 1 | 0.16% |
| How many children do you have? | 0 | 429 | 68.75% |
| | 1 | 92 | 14.74% |
| | 2 | 75 | 12.02% |
| | 3 or more | 17 | 2.72% |
| | Prefer not to answer | 11 | 1.76% |
| How many years of experience in software engineering do you have? | Less than 3 years | 222 | 35.58% |
| | More than 3, less than 5 years | 79 | 12.70% |
| | More than 5, less than 10 years | 125 | 20.00% |
| | More than 10, less than 15 years | 61 | 9.77% |
| | More than 15 years | 137 | 21.95% |
| What is your age range? | 18 - 24 | 82 | 13.14% |
| | 25 - 34 | 302 | 48.39% |
| | 35 - 44 | 141 | 22.59% |
| | 45 - 54 | 72 | 11.53% |
| | 55 - 64 | 25 | 4.00% |
| | Greater than 65 | 2 | 0.35% |

Finally, the majority (68.75%) of the sample reported not having children.

*B. RQ1: How does IP manifest in software professionals across different genders?*

A total of 52.72% of software professionals in the sample report experiencing frequent and intense levels of IP. The percentage suffering from IP per gender is illustrated in Figure 2. In this figure and the following ones, each bar is color-coded by gender and includes the percentage and sample size (N) for clarity. It is possible to observe that women and other genders are more frequently affected by IP than men.

In relation to race/ethnicity, Figure 3 illustrates its intersection with gender. In all intersectional figures, to avoid misleading representations, we only show groupings that had three or more participants. The number of Black women suffering from IP stands out. Interestingly, the Other race category is the only one where women report lower levels of IP than men. However, we emphasize that the sample sizes within the intersectional groupings are small, and while they provide some preliminary indications, more studies are necessary to have conclusive claims.

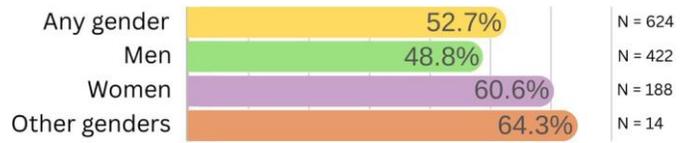

**Fig. 2:** Percentage of participants suffering from IP by gender.

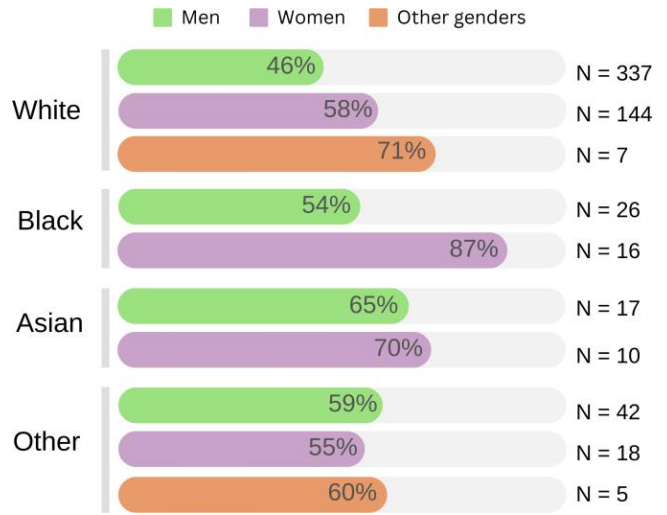

**Fig. 3:** Percentage of participants suffering from IP by race/ethnicity and gender.

With respect to marital status, our previous study [11] showed that IP is less prevalent among married individuals than among single individuals. In this paper, we analyze the intersection of gender and marital status. Figure 4 shows that the percentage of married participants suffering from IP was relatively lower than IP by non-married for all genders.

The study by Guenes *et al.* also indicated that parents experience lower levels of IP. The intersectional breakdown in Figure 5 shows that this reduction applies to men and women and that having two children reduces IP even further than having one.

One might hypothesize that married people and parents suffer less from IP because they tend to be older or more experienced. Indeed, Figure 6 indicates a general downward trend in the percentage suffering from IP with the increase of age for men, women, and all genders combined. Nonetheless, there seems to be an uptick observed when comparing the first age group of 18-24 to the subsequent 25-34 age group. Additionally, there is an upward trend for women starting at the age of 55, although it is important to note that the data for this age range is relatively sparse.

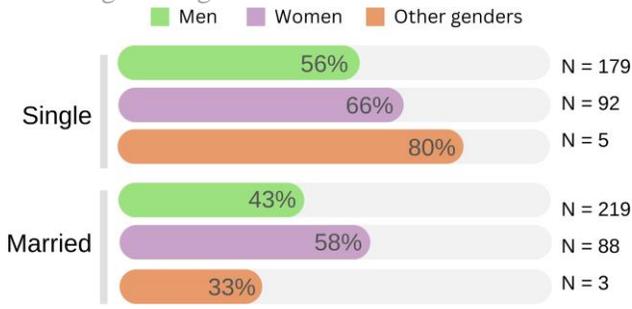

**Fig. 4:** Percentage of single and married participants suffering from IP by gender.

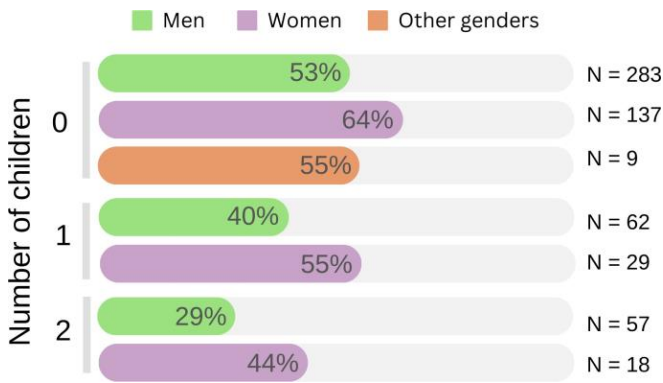

**Fig. 5:** Percentage of participants suffering from IP having 0, 1, or 2 kids by gender.

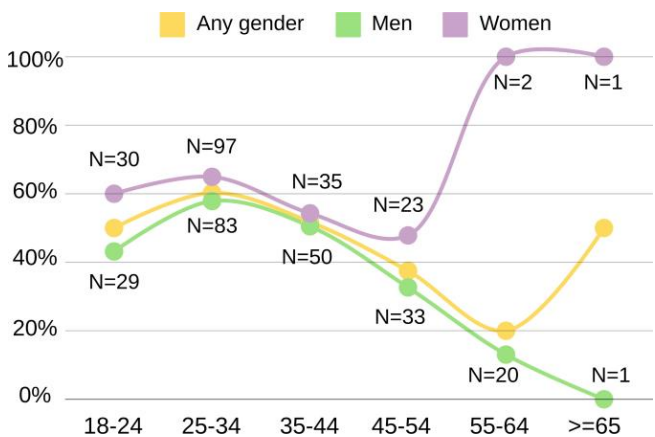

**Fig. 6:** Percentage of participants suffering from IP by age and gender.

Regarding the level of experience, Figure 7 reveals an unexpected pattern: while for men, women, and all genders combined, IP levels tend to decrease as experience increases, there is a noticeable slight resurgence of impostor feelings among individuals with 10–15 years of experience for all genders.

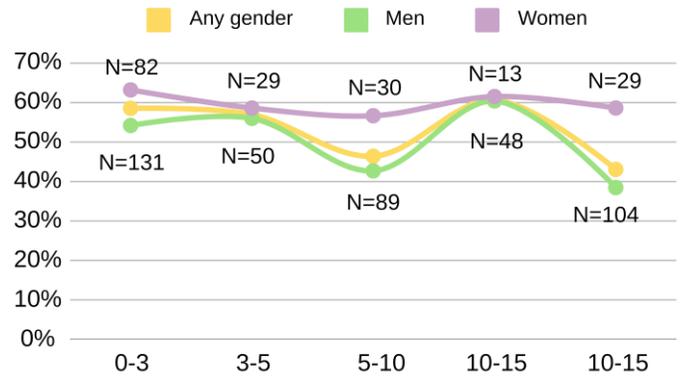

**Fig. 7:** Percentage of participants suffering from IP by years of experience and gender.

**Findings for RQ1:** Women experience IP more often than men, with a particularly high prevalence among black women, women who are single, and women without children. Age and experience appear to slightly reduce IP, particularly in men.

*C. RQ2: How does IP affect the well-being of software professionals across different genders?*

Answering RQ.2, Figure 8 shows the bootstrapped proportion of software engineering professionals with poor mental well-being (WHO-5 score between 0 and 50) by gender with the calculated confidence intervals. It reveals a worrying scenario in which more than 50% suffer from poor mental well-being (proportion of 58.33% confidence interval [54.49%, 62.50%]). It is possible to observe that large proportions of both men (56.87%, confidence interval [51.90%, 61.61%]) and women (61.17% confidence interval [53.72%, 6757%]) are seriously affected, with women experiencing it more frequently.

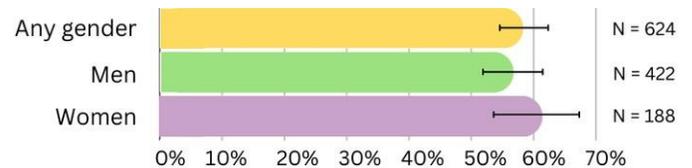

**Fig. 8:** Percentage of software professionals with poor mental well-being (WHO-5 score from 0 to 50) and confidence intervals, by gender.

Having characterized well-being, to answer RQ2, we wanted to understand how well-being scores vary for people who suffer from IP or not. Therefore, the boxplot in Figure 9 shows the WHO-5 well-being scores for all genders, men, and women suffering from IP (left) and not suffering from IP (right). For individuals with IP, both women and men exhibit lower median WHO-5 scores compared to those without IP.

Among those with IP, the first quartile (Q1), median, and third quartile (Q3) scores for women and men are remarkably close. This similarity indicates that individuals experiencing IP, regardless of gender, report similarly low levels of subjective well-being. For those without IP, the scores are also closely aligned across gender groups.

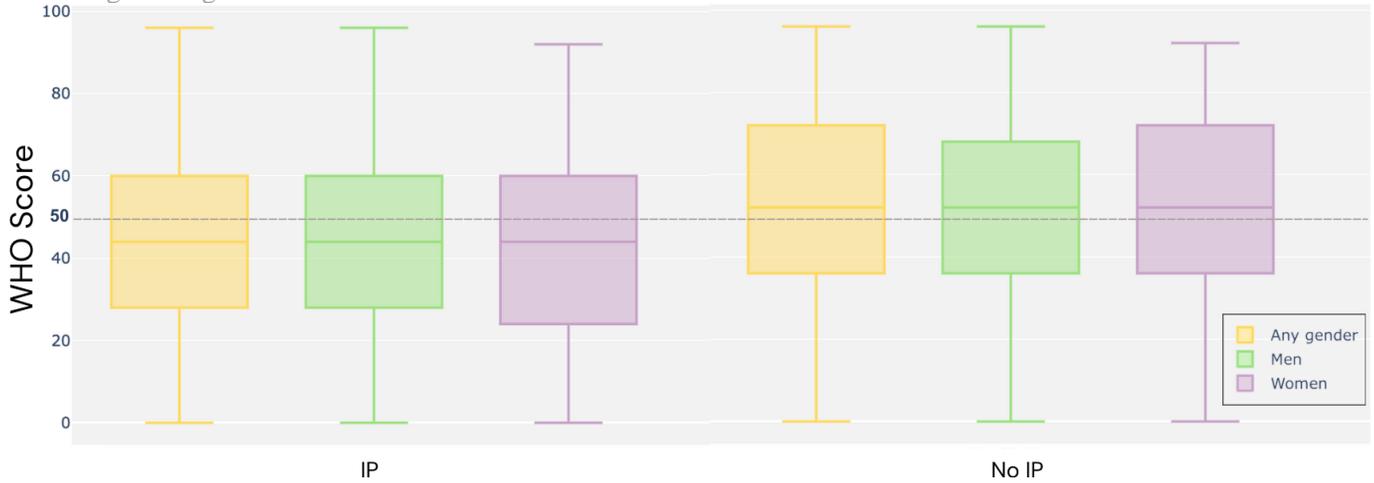

Fig. 9: WHO-5 well-being scores for all genders, men, and women, grouped by suffering from IP (left side) or not (right side).

Notably, the median scores for both women and men experiencing IP fall below the 50 threshold, indicating poor mental well-being. In contrast, the median scores for women and men who do not experience IP are above the 50 threshold, reflecting adequate or good mental well-being. Still, we wanted to verify if the observed differences were statistically significant.

When comparing WHO-5 index scores for software professionals with frequent and intense IP symptoms with those with few or moderate symptoms, we found a statistically significant difference between the groups. This indicates that on the WHO-5 well-being index, software professionals with IP consistently show significantly lower subjective well-being than their counterparts without IP. Men and women with IP also show significantly lower subjective well-being on the WHO-5 index compared to men and women without IP (Mann-Whitney U-test $p$-value $<$ 2.49e-11 for men and $p$-value $<$ 4.61e-05 for women).

To allow a better understanding of this statistical difference confirmed for the population, we show it graphically in terms of frequencies of people with poor mental well-being within our sample. Figure 10 shows the percentage and the number of participants in the sample with poor mental well-being (WHO-5 score 0-50) per gender, grouped by suffering or not from IP. It is easy to observe that within the groups that do not suffer from IP, for all genders, the percentage of people with poor mental well-being is much lower.

Figure 11 shows the frequencies of people who fall within the threshold associated with major depression (WHO-5 score 0-28). For this more stringent threshold, the differences between the groups that do and do not suffer from IP are much larger, approximately doubling the percentage for both, men and women.

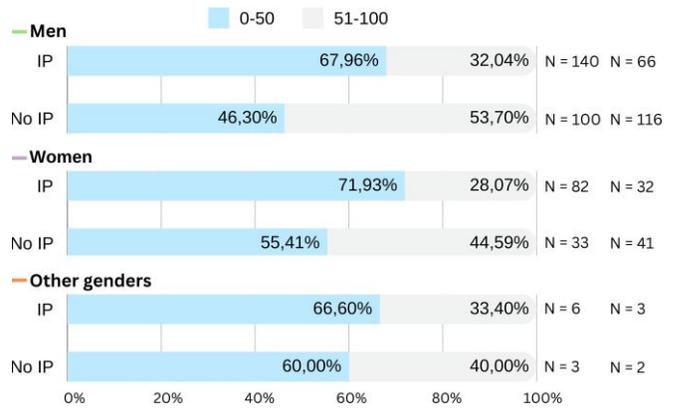

Fig. 10: Percentage of participants with low levels of well-being per gender identities, grouped by whether they suffer from IP or not.

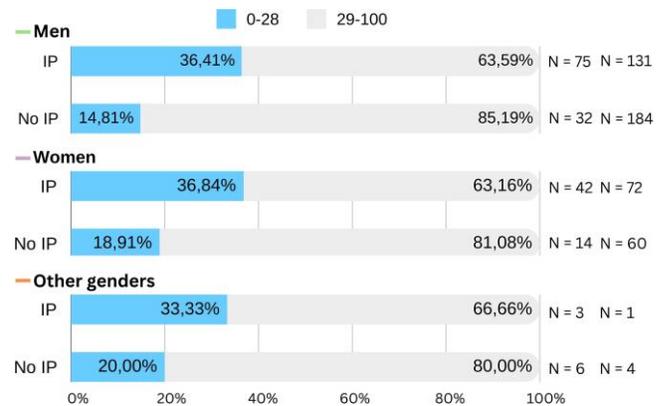

Fig. 11: Percentage of participants with levels of well-being of major depression per gender identities, grouped by whether they suffer from IP or not.

> **Findings for RQ2:** A significant proportion of software engineering professionals experience poor mental well-being, with women more frequently affected. IP is strongly linked to poor mental health, with individuals experiencing IP showing significantly lower well-being and a higher prevalence of scores associated with depression, across all genders.

## V. Discussion

We investigated how IP manifests among software engineering professionals, focusing on intersections of gender with race/ethnicity, marital status, number of children, age, and software engineering experience. Additionally, we explored the well-being of those experiencing IP to provide insights into how these factors interplay.

Our results align with the high impact IP has on women [23], revealing that women experience IP more frequently and intensely than men. Notably, our study also included individuals with other gender identities, who were found to experience IP as much as women or even more. This finding suggests that marginalized gender groups may be disproportionately affected by systemic and cultural factors contributing to IP.

One of the findings that caught our attention is the intersection between gender and race/ethnicity. Black women showed a particularly high frequency of experiencing IP (87%). Historical marginalization and systemic exclusion in the computing field likely play a significant role in this disparity [2]. Contributions by black women, such as those who programmed mainframes for NASA's space program, had been erased from computing history [20]. This erasure reinforces false narratives, such as the idea that black women lack interest in computer science or that white men naturally excel at programming, which helps to increase IP and perpetuate stereotypes in the field.

When examining age and professional experience, our findings suggest that these factors influence IP in complex ways. Early career professionals (less than three years of experience) may face pressure to demonstrate their knowledge and meet elevated social and professional expectations. In contrast, more experienced professionals may experience IP resurfacing due to increased visibility, higher expectations, and greater responsibilities within their organizations. This aligns with the definition of IP as a fear of being exposed as a fraud among high-achieving individuals [5].

In terms of well-being, large proportions of software engineering professionals showed poor mental well-being, with women being more frequently affected. Both, men and women experiencing IP showed statistically significant lower well-being scores compared to their counterparts without IP, with consistent patterns observed across genders. This finding aligns with previous research that showed a similar impact of burnout in attrition across genders [24], also suggesting that IP negatively affects subjective well-being regardless of gender. In particular, people with frequent and intense symptoms of IP were more likely to report scores associated with depression. These results strengthen the link between IP and well-being, heling raise awareness of the general importance of the subject.

In our previous work [11], we revealed the high prevalence of IP among software professionals and its significant impact on perceived productivity. This research expands our findings by incorporating intersectional gender analyses and examining the relationship between IP and well-being. Uncovering the prevalence of IP across diverse gender intersections and providing evidence of its detrimental effects on mental health, this study contributes to a better understanding of how IP manifests within the software engineering profession. These insights have implications for the development of interventions aimed at addressing IP and improving the mental health of software professionals. Specifically, by identifying the most affected demographic groups, such as women, particularly black women, and people with marginalized gender identities, targeted strategies can be designed to reduce IP and its associated mental health challenges. Furthermore, we believe that addressing systemic issues such as representation and inclusion in the computing field is critical to reducing the underlying factors that contribute to IP.

## VI. Threats to Validity and Limitations

All threats to validity typically considered for survey research [13], including face validity, content validity, criterion validity, and reliability, were thoroughly discussed in our prior paper by Guenes *et al.* [11]. Since this study builds upon our previous research by utilizing our data and analyzing new constructs, the threat analysis from the previous paper is referenced, while the current paper addresses potential threats related to the new analysis.

Construct validity. The data was collected using psychometrically validated instruments for IP (CIPS scale) and well-being (WHO-5 index). Both CIPS and the WHO-5 index are widely used and reliable instruments [14], [15].

Conclusion Validity. Part of our study consisted of intersectional analyses for different genders. Some of these intersections had a limited number of data points, which hindered the applicability of inferential statistics. Furthermore, the WHO-5 index is an ordinal scale. Therefore, we applied bootstrapping only on the proportion of professionals scoring below a certain threshold and not on the WHO-5 score itself. In addition, we conservatively tested the statistical significance of the relationship between IP and well-being using the Mann-Whitney U test, which is appropriate for ordinal scales.

External Validity. The sample size and representativeness evaluations conducted previously (Guenes *et al.* [11]) increase our confidence in generalizing the results for which we applied inferential statistics. These results include the bootstrapped proportion of software engineering professionals with low levels of well-being and the statistically significant relations between IP and well-being. For intersections with limited data points, we only reported frequencies within the sample to provide some preliminary indications without external validity claims.

## VII. Conclusion

This study examined how IP manifests among software professionals, considering gender, race/ethnicity, marital status, number of children, age, and experience. It also explored the relationship between IP and mental well-being. Findings show that IP is more prevalent among women, particularly Black women, single women, and women without children. While age and experience may slightly reduce its prevalence, no conclusive patterns were observed. Additionally, the study revealed a strong link between IP and mental health, with affected professionals exhibiting significantly lower well-being and a higher prevalence of depression-related symptoms.

Our findings contribute to a better understanding of how IP manifests within intersectional gender demographics and its detrimental effects on well-being. They highlight the need for targeted interventions and support mechanisms to address IP, particularly for groups disproportionately affected.

Future research should focus on identifying systemic causes of IP disparities and developing targeted interventions to mitigate its effects, particularly in software engineering. Investigating strategies from other fields and adapting them to the unique challenges of the software industry could provide valuable insights. Additionally, a deeper and more nuanced understanding of IP can be achieved by combining broad quantitative analyses with qualitative perspectives that capture individual experiences. Future work should also explore how organizational factors, such as company culture, leadership styles, and diversity, equity, and inclusion (DEI) initiatives, influence the prevalence and intensity of IP. Empirical assessments of these strategies would help establish recommendations for companies to support professionals facing IP, fostering psychologically safe environments that enhance well-being, productivity, and workplace mental health.